\documentclass[12pt,14paper]{article}
\usepackage{latexsym}
\usepackage{amssymb}
\usepackage{amsmath}
\usepackage{amsfonts}
\usepackage{amsopn}
\usepackage{graphicx}
\usepackage{gastex}

\newtheorem{theorem}{Theorem}
\newtheorem{lemma}{Lemma}

\newtheorem{example}{Example}

\newtheorem{definition}{Definition}
\begin{document}

\title{On automatic infinite permutations}
\author{A. Frid, L. Zamboni}
                       
\maketitle
\begin{abstract}
An infinite permutation $\alpha$ is a linear ordering of $\mathbb N$.
We study properties of infinite permutations analogous to those of infinite words,
and show some resemblances and some differences between permutations and words.
In this paper, we try to extend to permutations the notion of automaticity. As we shall show, the standard definitions which are equivalent in the case of words are not equivalent in the context of permutations. We investigate the relationships between these definitions and prove that they constitute a chain of inclusions. We also construct and study an automaton generating the Thue-Morse permutation.

Une permutation infinie est un ordre total sur $\mathbb N$. Nous \'etudions les caract\'eristiques des permutations infinies qui sont analogues \`a celles des mots infinis, et nous montrons que certaines d'entre elles se comportent de la m\^eme fa\c con, et d'autres pas. Dans cet article nous essayons d'\'etendre la notion d'automaticit\'e de mots aux permutations. Cependant il arrive que des d\'efinitions \'equivalentes pour les mots ne sont pas \'equivalentes pour les permutations. Nous \'etudions la relation entre ces d\'efinitions en d\'emontrant qu'ils constituent une cha\^ine d'inclusions. En outre, nous discutons les automates engendrant la permutation de Thue-Morse.
\end{abstract}

\section{Infinite permutations}
Let $S$ be a finite or countable ordered set: we shall typically take $S$ to be 
either the set $\mathbb N=\{0,1,2,\ldots\}$ of all non-negative integers, or  some its subset.
Let ${\mathcal A}_S$ be the set of all sequences of pairwise distinct reals indexed by
the set  $S$. Define an equivalence relation $\sim$ on ${\mathcal A}_S$
as follows: given $a,b$ in${\mathcal A}_S$, with $a=\{a_s\}_{s\in S}$ and
$b=\{b_s\}_{s\in S}$; we write $a \sim b$ if and only if for all $s,r
\in S$ the inequalities $a_s < a_r$ and $b_s<b_r$ hold or do not
hold simultaneously. An equivalence class from ${\mathcal A}_S / \sim$ is
called an {\it \mbox{($S$-)}permutation}. If an $S$-permutation
$\alpha$ is realized by a sequence of reals $a$, that is, if the sequence $a$ belongs to the class $\alpha$, we
write $\alpha=\overline{a}$. In particular, a
$\{1,\ldots,n\}$-permutation always has a representative with all
values in $\{1,\ldots,n\}$, i.~e., can be identified with a usual
permutation in $S_n$.

In equivalent terms, a permutation can be considered as a linear ordering of $S$
which may differ from the ``natural'' one. That is, for $i,j \in S$, the natural
order between them corresponds to $i<j$ or $i>j$, while the ordering we intend to
define corresponds to $\alpha_i <\alpha_j$ or $\alpha_i >\alpha_j$. We shall also
use the symbols $\gamma_{ij}\in \{<,>\}$ meaning the relations between $\alpha_i$
and $\alpha_j$, so that by definition we have $\alpha_i \gamma_{ij} \alpha_j$ for
all $i \neq j$.

\begin{example}{\rm
Let $\{a_i\}_{i=0}^{\infty}$ be the sequence defined by $a_n=(-1/2)^n$, and $\{b_i\}_{i=0}^{\infty}$ be the sequence defined by 
$b_i=1000+(-1)^n/n$. Then $\overline{a}=\overline{b}$; and we also can define the respective permutation $\alpha=\overline{a}=\overline{b}$ directly by the family of inequalities: for all $i,j\geq 0$ we have $\alpha_{2i}>\alpha_{2j+1}$, $\alpha_{2i}>\alpha_{2i+2}$, and $\alpha_{2j+1}<\alpha_{2j+3}$. Equivalently, the same family of inequalities can be written as $\gamma_{2i,2j+1}=>$, $\gamma_{2i,2i+2}=>$, and $\gamma_{2j+1,2j+3}=<$. It can be easily checked that these inequalities completely define the permutation, and that it is equal to $\overline{a}$ and to $\overline{b}$.

Note also that the permutation $\alpha$ cannot be represented by a sequence of integers since $\alpha_1<\alpha_n< \alpha_0$ for all $n\geq 2.$ 
}
\end{example}

For more background on the theory of infinite permutations, we refer the reader to \cite{degs}. Periodicity, subword complexity and maximal pattern complexity of permutations were studied in \cite{ff,afks}.

Any aperiodic (non ultimately periodic) infinite word $w=w_0 w_1 w_2 \cdots w_n \cdots$ 
on a finite alphabet $\Sigma_q=\{0,\ldots,q-1\}$ naturally defines an infinite permutation $\alpha$ represented by the sequence of reals $\{a_i\}_{i=0}^\infty$ defined by $a_i=.w_i w_{i+1} \cdots=\sum_{j=0}^{\infty}\frac{w_{i+j}}{q^{j+1}}.$ Such a permutation is said to be {\em valid} over the alphabet $\Sigma_q$. Valid permutations have been investigated by Makarov \cite{makarov,mak_pd,mak_tm,mak2}. It is not difficult to see that there exist infinite permutations which are not valid, including for example the monotonic ones. 

\begin{example} \label{tm}
{\rm 
Let $w_{TM}$ be the Thue-Morse word, $w_{TM}=w_0w_1w_2 \cdots=01101001\cdots$: here $w_i$ is the parity of 1s in the binary representation of $i$. Then the associated infinite permutation $\alpha_{TM}$  is the order among the binary numbers $.01101001\cdots$, $.1101001\cdots$, $.101001\cdots$, $.01001\cdots$ The first four values are ordered as $\alpha_3<\alpha_0<\alpha_2 < \alpha_1$. In terms of the symbols $\gamma_{ij}\in \{<,>\}$ we have $\gamma_{01}=\gamma_{02}=<$ and $\gamma_{03}=\gamma_{12}=\gamma_{13}=\gamma_{23}=>$, etc.  The Thue-Morse permutation has been considered in detail in \cite{mak_tm,widmer}.
}
\end{example}
\section{Automatic words and permutations}
We begin by recalling some of the basic notions concerning automatic words. For more background on this topic we refer the reader to the book by Allouche and Shallit \cite{ash} where many details and examples can be found.

Let $k>1$ be a positive integer. An infinite word $w=w_0 w_1 w_2\cdots$ over $\Sigma_q$ is called {\em $k$-automatic} if its $n$th symbol $w_n$ is the output of a deterministic finite automaton after feeding to it the base $k$ representation $(n)_k$ of $n$. Formally, we define the automaton 
$A=(Q,\Sigma_k,\delta,q_0,\Sigma_q,\tau)$ with $\delta: Q \times \Sigma_k \to Q$ (and the natural extension of $\delta$  to a function $Q \times \Sigma^*_k \to Q$) and $\tau: Q \to \Sigma_q$ so that $w_n=\tau(\delta(q_0,(n)_k))$ for all $n\geq 0$.

\begin{example}
{\rm 
The Thue-Morse word $w_{TM}=w_0 w_1 \cdots = 01101001\cdots$ is 2-automatic by the definition given in the previous example. The corresponding automaton is depicted below.

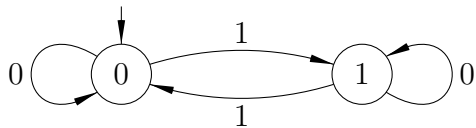
\begin{center}
\unitlength 0.8mm
\begin{picture}(80,20)(-10,0)
\gasset{AHLength=4.0,AHlength=4,AHangle=9.09}
\gasset{Nw=10,Nh=10,Nmr=5}
\gasset{ELside=l}\gasset{loopCW=n}
  \node(N1)(10,10){$0$}
  \node(N2)(50,10){$1$}
  \drawedge[curvedepth=4](N1,N2){$1$}
  \drawedge[curvedepth=4](N2,N1){$1$}
  \drawloop[loopangle=180,ELside=r,ELpos=50](N1){$0$}
  \drawloop[loopangle=0,ELside=r,ELpos=50](N2){$0$}
  \imark[iangle=90](N1)
\end{picture}
\end{center}
}
\end{example}

There are several well known equivalent definitions of automatic words (see \cite{ash} for a more detailed discussion). One such alternative definition uses uniform morphisms.

A {\em morphism} $\varphi: \Delta^* \to \Sigma^*$, where $\Delta$ and $\Sigma$ are alphabets, is a mapping satisfying $\varphi(xy)=\varphi(x)\varphi(y)$ for all $x,y$. Clearly, a morphism is completely determined by the images of letters. A morphism is called {\em $k$-uniform} if the image of each letter is of length $k$. A {\em fixed point} of a uniform morphism $\varphi: \Delta \to \Delta^k$ is a (right) infinite word $w$ satisfying $w=\varphi(w)$; a fixed point of $\varphi$ always starts with a letter $a$ such that $\varphi(a)$ starts with $a$. 

A 1-uniform morphism $c: \Delta \to \Sigma$ is called a {\em coding}.

\begin{theorem}[Cobham,\cite{cobham,ash}]
For each $k>1$, an infinite word $w$ is $k$-automatic if and only if it is the image under a coding of a fixed point of a 
$k$-uniform morphism. 
\end{theorem}

\begin{example}{\rm
The Thue-Morse word is the fixed point of the 2-uniform morphism $\varphi: 0 \mapsto 01, 1 \mapsto 10$. The coding $c$ is here trivial. }
\end{example}

For another equivalent definition, we define the {\em $k$-kernel}  of an infinite word $w=w_0 w_1 \cdots$ to be the set of arithmetic subsequences of $w$ of the form
$w_{i} w_{k^n+i} w_{2k^n+i}\cdots$ for some $n\geq 0$ and $0\leq i<k^n$.

\begin{theorem}[Eilenberg,\cite{eil,ash}]
For each $k>1$, an infinite word $w$ is $k$-automatic if and only if its $k$-kernel is finite.
\end{theorem}

\begin{example}{\rm
The 2-kernel of the Thue-Morse word contains just two elements: the Thue-Morse word itself and the word obtained from it by exchanging 0s and 1s. }
\end{example}

We now consider analogues of the previous three definitions  of automatic words in the context of infinite permutations\footnote{Yet another widely used equivalent definition of $k$-automatic words involves algebraic formal power series \cite{christol,ash}. However, we do not consider  formal power series in the context of  permutations.}. The first unfortunately only applies to valid permutations.

\begin{definition}{\rm
A valid permutation is {\em $V$-$k$-automatic} if it is generated by a $k$-automatic word over a finite alphabet. The class of all $V$-$k$-automatic permutations is denoted by $\mathcal V_k$.
}
\end{definition}
\begin{example}{\rm
The Thue-Morse permutation $\alpha_{TM}$ from Example \ref{tm} is $V$-$2$-automatic since the Thue-Morse word is 2-automatic.}
\end{example}

Our next definition directly involves  an automaton, and so applies more generally:

\begin{definition}{\rm
A permutation $\alpha=\alpha_0 \alpha_1 \cdots$ is {\em $A$-$k$-automatic} if there exists a deterministic finite automaton
$\mathbb A=(Q,(\Sigma_k)^2,\delta,q_0,\{<,>,=\},\tau)$ with $\delta: Q \times (\Sigma_k)^2 \to Q$ (and the natural extension of $\delta$  to a function $Q \times (\Sigma_k^2)^* \to Q$) and $\tau: Q \to \{<,>,=\}$ so that $\gamma_{ij}=\tau(\delta(q_0,(i)_k\times (j)_k))$ for all $i,j \geq 0$.
The class of all $A$-$k$-automatic permutations is denoted by $\mathcal A_k$.
}
\end{definition}
According to this definition, the automaton $\mathbb A$ is fed by pairs of digits from the base $k$ representations of $i$ and $j$ (passing the automaton simultaneously, starting with the most significant digit or the starting 0 if necessary). The output is the relation between the elements of $\alpha$ numbered $i$ and $j$.

Note that not all the automata of this form define permutations. However, in practice it is not too difficult to check whether the automaton in question actually  generates a permutation, that is, an order on $\mathbb N$.

\begin{lemma}
Given an automaton $\mathbb A=(Q,(\Sigma_k)^2,\delta,q_0,\{<,>,=\},\tau)$, it is decidable if it generates a permutation or not.
\end{lemma}
{\sc Proof.} We must check that the relation constructed is antisymmetric and transitive, that is, that it is an order. To check the antisymmetric property, consider the square automaton 
$$\mathbb A^2=(Q^2,((\Sigma_k)^2)^2,\delta',q_0 \times q_0,\{<,>,=\}^2,\tau)$$
where the transition function $\delta': Q^2 \times ((\Sigma_k)^2)^2 \to Q^2$ 
is defined by $\delta'(q_1\times q_2, (i_1,j_1)\times (i_2,j_2)) \to \delta(q_1, (i_1,j_1)) \times \delta(q_2, (i_2,j_2))$, 
and $\tau(q_1\times q_2)=(\tau(q_1),\tau(q_2))$. Now consider the restriction of $\mathbb A^2$ to the input of the form $(i,j)\times (j,i)$. All the reachable states must give the output $(<,>)$, $(>,<)$, or $(=,=)$; moreover, the states giving $(=,=)$ must be reachable only by the transitions of the form $(i,i),(i,i)$, and all the other reachable states cannot be reached by the input of that form. Clearly, this property can be checked by standard means.

To check the transitive property, we analogously consider the cube automaton $\mathbb A^3$ and its restriction to the input of the form $(i,j)\times (j,k) \times (k,i)$. In this case, the reachable states of this subautomaton should not give the output $(<,<,<)$ nor $(>,>,>)$. \hfill $\Box$

Note also that due to this definition, an  $A$-$k$-automatic permutation is equivalent to a very specific two-dimensional $[k,k]$-automatic word with entries $\gamma_{ij}$ over the alphabet $\{<,>,=\}$. For the properties of two-dimensional automatic words, see Chapter 14 of \cite{ash}.

Our third and last definition of $k$-automatic permutations involves  $k$-kernels. Similarly to words, let us define the $k$-kernel of a permutation $\alpha=\alpha_0 \alpha_1 \cdots$ as the set of all permutations of the form $\alpha_{i} \alpha_{k^n+i} \alpha_{2k^n+i}\cdots$ for some $n\geq 0$ and $0\leq i<k^n$ (interpreted as $\mathbb N$-permutations).

\begin{definition}{\rm
A permutation is {\em $K$-$k$-automatic} if its $k$-kernel is finite. The class of all $K$-$k$-automatic permutations is denoted by $\mathcal K_k$.
}
\end{definition}

The main result of the paper is the following

\begin{theorem}
For each $k \geq 2$, we have $\mathcal V_k \subsetneq \mathcal A_k \subsetneq \mathcal K_k$. 
\end{theorem}
Thus, it seems that no equivalence similar to that for words is possible for infinite permutations. 

In particular, it follows from the theorem that the Thue-Morse permutation is $A$-2-automatic and $K$-2-automatic. The latter fact is easy to check since once again there are only two elements in the 2-kernel. An automaton defining the Thue-Morse permutation is shown below.

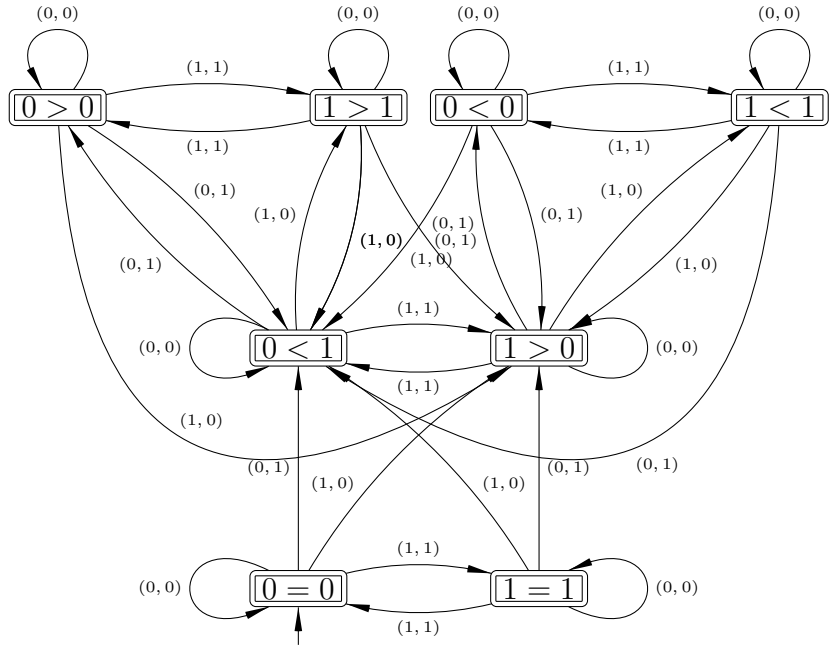
\begin{center}
\unitlength 0.8mm
\begin{picture}(80,110)(-10,0)
\gasset{AHLength=4.0,AHlength=4,AHangle=9.09}
\gasset{Nw=10,Nh=10,Nmr=5}
\gasset{ELside=l}\gasset{loopCW=n}
\gasset{Nadjust=w,Nadjustdist=2,Nh=6,Nmr=1}
  \node(N1)(10,10){$0=0$}
  \node(N2)(50,10){$1=1$}
  \node(N3)(10,50){$0<1$}
  \node(N4)(50,50){$1>0$}
  \node(N5)(-30,90){$0>0$}
  \node(N6)(20,90){$1>1$}
  \node(N7)(40,90){$0<0$}
  \node(N8)(90,90){$1<1$}
  {\tiny
  \drawedge[curvedepth=4](N1,N2){$(1,1)$}
  \drawedge[curvedepth=4](N2,N1){$(1,1)$}
  \drawedge[curvedepth=4](N3,N4){$(1,1)$}
  \drawedge[curvedepth=4](N4,N3){$(1,1)$}
  \drawedge[curvedepth=4,ELpos=30](N1,N4){$(1,0)$}
  \drawedge[curvedepth=-4,ELside=r,ELpos=30](N2,N3){$(1,0)$}
  \drawloop[loopangle=180,ELside=r,ELpos=50](N1){$(0,0)$}
  \drawloop[loopangle=0,ELside=r,ELpos=50](N2){$(0,0)$}
  \drawloop[loopangle=180,ELside=r,ELpos=50](N3){$(0,0)$}
  \drawloop[loopangle=0,ELside=r,ELpos=50](N4){$(0,0)$}
  \drawedge(N1,N3){$(0,1)$}
  \drawedge[ELside=r](N2,N4){$(0,1)$}
  \imark[iangle=-90](N1)
  \rmark(N1)
\rmark(N2)
\rmark(N3)
\rmark(N4)

 \rmark(N5)
\rmark(N6)
\rmark(N7)
\rmark(N8)
 \drawloop[loopangle=90,ELside=r,ELpos=50](N5){$(0,0)$}
  \drawloop[loopangle=90,ELside=r,ELpos=50](N8){$(0,0)$}
  \drawloop[loopangle=90,ELside=r,ELpos=50](N6){$(0,0)$}
  \drawloop[loopangle=90,ELside=r,ELpos=50](N7){$(0,0)$}
 \drawedge[curvedepth=4](N5,N6){$(1,1)$}
  \drawedge[curvedepth=4](N6,N5){$(1,1)$}
   \drawedge[curvedepth=4](N7,N8){$(1,1)$}
  \drawedge[curvedepth=4](N8,N7){$(1,1)$}
  \drawedge[curvedepth=4](N3,N5){$(0,1)$}
  \drawedge[curvedepth=4](N5,N3){$(0,1)$}
   \drawedge[curvedepth=4](N3,N6){$(1,0)$}
  \drawedge[curvedepth=4](N6,N3){$(1,0)$}
    \drawedge[curvedepth=4](N4,N7){$(0,1)$}
  \drawedge[curvedepth=4](N7,N4){$(0,1)$}
   \drawedge[curvedepth=4](N4,N8){$(1,0)$}
  \drawedge[curvedepth=4](N8,N4){$(1,0)$}
   \drawedge[curvedepth=4](N6,N3){$(1,0)$}
    \drawedge[curvedepth=-40](N5,N4){$(1,0)$}
  \drawedge[curvedepth=-4](N6,N4){$(0,1)$}
   \drawedge[curvedepth=4](N7,N3){$(1,0)$}
  \drawedge[curvedepth=40](N8,N3){$(0,1)$}
}
\end{picture}
\end{center}

The vertices of the automaton are labeled with the respective symbols $w_i$ and $w_j$ of the Thue-Morse word and the relation between $.w_iw_{i+1}\cdots$ and $.w_jw_{j+1}\cdots$.

Note that the subautomaton in the lowest row corresponds to the trivial situation of $i=j$ and is isomorphic to the usual Thue-Morse automaton. Moreover, there are no edges incoming to this subautomaton from the outer vertices.

\section{Proof of the inclusions}
We begin with the simpler aspects of our proof.
\begin{lemma}
For all $k>1$ we have $\mathcal A_k \backslash \mathcal V_k \neq \emptyset$.
\end{lemma} 
{\sc Proof.} The monotonic permutation $\alpha$ with just $\alpha_i < \alpha_{i+1}$ for all $i$ belongs to $\mathcal A_k \backslash \mathcal V_k$ since it can be constructed by a trivial automaton but is not valid, as discussed earlier. \hfill $\Box$

\begin{lemma}
For all $k>1$ we have $\mathcal K_k \backslash \mathcal A_k \neq \emptyset$.
\end{lemma} 
{\sc Proof.} We construct a permutation $\alpha$ from $\mathcal K_2 \backslash \mathcal A_2$ as follows: Let us state that for all $j$ we have $\alpha_j<\alpha_{j+2}$ and $\alpha_{2j+1}<\alpha_{2j+2}$. Finally, let us fix a binary
word $u=u_0u_1\cdots$ over the alphabet $\{<,>\}$ which is {\em not} 2-automatic and define the relation $\gamma_{2j,2j+1}$ between $\alpha_{2j}$ and $\alpha_{2j+1}$ to be equal to $u_j$, so that $\alpha_{2j} u_j \alpha_{2j+1}$. Then the 2-kernel of $\alpha$ is of cardinality two: it just contains $\alpha$ itself and the monotonically increasing permutation. So, $\alpha$ is $K$-$2$-automatic. On the other hand, suppose that it is $A$-2-automatic. Then from the automaton determining $\gamma_{ij}$ from the binary representations of $i$ and $j$, the automaton determining the sequence of $\gamma_{2j,2j+1}=u_j$ from the binary representation of $j$ could be derived by a standard procedure. But this automaton does not exist, a contradiction.

Examples for greater values of $k$ may be constructed analogously: we simply assume that all the elements of the $k$-kernel except for the permutation itself are monotonic, and define the relations between neighbouring entries of the permutation in a complicated fashion. \hfill $\Box$.

\begin{lemma}
For all $k>1$ we have $\mathcal A_k \subseteq \mathcal K_k$.
\end{lemma} 
{\sc Proof.} We note that a permutation $\alpha$ can be interpreted as a specific two-dimensional word $(\gamma_{ij})_{i,j=0}^\infty$, and by the definition, $\alpha$ is $A$-$k$-automatic if and only if that two-dimensional word is $[k,k]$-automatic. The $k$-kernel of $\alpha$ also corresponds to the $[k,k]$-kernel of that word, which is finite (see Theorem 14.2.2 in \cite{ash}). \hfill $\Box$

Now let us prove the least trivial part of the result.
\begin{lemma}
For all $k>1$ we have $\mathcal V_k \subseteq \mathcal A_k$.
\end{lemma} 
{\sc Proof.} Let us consider a $k$-automatic word $v$ generating a valid permutation $\alpha$, the $k$-uniform morphism $\varphi$ and the coding $c$ such that $v=c(w)$, where the infinite word $w=\varphi(w)=w_0w_1\cdots$ over a finite alphabet $\Delta$ of cardinality $d$ is a fixed point of $\varphi: \Delta \to \Delta^k$. We shall use $\varphi$ and $c$ to construct directly the automaton $\mathbb A =(Q,(\Sigma_k)^2,\delta,q_0,\{<,>,=\},\tau)$.

In what follows for all $n\geq 0$ we shall use the notation $T^{n}w$ for the shift $w_n w_{n+1} \cdots$ of the sequence $w$. For finite or infinite words $u'=u'_0 u'_1\cdots$ over an alphabet $\Delta'$ and $u''=u''_0 u''_1\cdots$ over $\Delta''$ we shall use the notation $u' \times u''$ for the word $(u'_0\times u''_0) (u'_1\times u''_1) \cdots$ over the alphabet $\Delta'\times \Delta''$.

Let us denote the set of all factors of $w$ of length 2 by $P$, so that $P \subseteq \Delta^2$. The number $p$ of such factors is not greater than $d^2$. 
Now consider the product $P\times P$ and denote by $S_{P \times P}$ the set of all permutations of the elements of $P\times P$ and the new symbol $\diamond$ which is a marker. Thus the number of such permutations is equal to $(p^2+1)! \leq (d^4+1)!$.

The set $S_{P\times P}$ is the set of the states of the automaton $\mathbb A$. Denote by $[a]_k$ the integer whose $k$-ary representation is the string $a \in \Sigma_k^*$. Then the state corresponding to the input $a \times b \in (\Sigma^2_k)^*$ is the following: first, order all the factors of length 2 of 
$T^{[a]_k}w \times T^{[b]_k}w$ in order of appearance; then $\diamond$; then all the remaining words of $P \times P$ in any fixed order (say, in the lexicographic order).

The starting state $q_0$ corresponds to the input $0 \times 0$ and thus is equal to $(a_1\times a_1,a_2\times a_2,\ldots , a_p \times a_p,\diamond,\ldots)$. Here $a_1,\ldots,a_p$ are the factors of $w$ of length 2 in order of appearance: we count overlapping factors as well, that is, we take the sequence $w_0w_1$, $w_1w_2$, $w_2w_3$, etc., and erase all words which we have met before. The final dots indicate all the other elements of $P \times P$ arranged in lexicographic order.

Now let us define the transition function $\delta$. Given a state $q=(s_1t_1\times p_1r_1,\ldots,s_lt_l\times p_lr_l,\diamond,\ldots)$, where $s_m,t_m,p_m,r_m \in \Delta$ for all $m$, and a pair $i \times j$, where $0\leq i,j <k$, we define the state $\delta(q,i\times j)$ as follows.

First for each $0\leq i,j <k$ let us define a function $f_{ij}: P\times P \to (P \times P)^k$ as follows. Let $st\times pr\in P\times P$; consider $\varphi(st)=g_0\cdots g_{2k-1}$ and $\varphi(pr)=h_0\cdots h_{2k-1}$. Then $f_{ij}(st \times pr)= (g_ig_{i+1} \times h_j h_{j+1}, g_{i+1}g_{i+2} \times h_{j+1} h_{j+2},\ldots, g_{i+k-1}g_{i+k} \times h_{j+k-1} h_{j+k})$.

Now to define $\delta(q,i\times j)$ we write down successively the elements of $f_{ij}(s_1t_1\times p_1r_1)$, $\ldots$, $f_{ij}(s_lt_l\times p_lr_l)$, and then read them from left to right deleting the elements which have appeared in the string before. The resulting sequence of elements of $P \times P$ is the part of $\delta(q,i\times j)$ preceding the diamond, so that it remains to complete it with $\diamond$ and then by all the other elements of $P \times P$ in the lexicographic order.

Let us show that if a state $q$ describes the order of elements of length 2 of $T^{[a]_k}w \times T^{[b]_k}w$ for some $a\times b \in (\Sigma^2_k)^*$, then the state $\delta(q,i\times j)$ does describe the order of elements of length 2 of $T^{[ai]_k}w \times T^{[bj]_k}w$. In fact, it is evident from the construction that if  $s_it_i$ appears for the first time at the position numbered $n$ of $T^{[a]_k}w$, then $\varphi(s_it_i)=g_0\cdots g_{2k-1}$ appears in $\varphi(T^{[a]_k}w)=T^{[a0]_k}w$ at the position numbered $kn$. For $T^{[b0]_k}w$ we can make the analogous statement; so, $g_i \cdots g_{i+k} \times h_j \cdots h_{j+k}$ really appears in $T^i(T^{[a0]_k}w) \times T^j(T^{[j0]_k}w) =  T^{[ai]_k}w \times T^{[bj]_k}w$ at the position numbered $kn$. Now our procedure just considers successively the $k$ factors of length 2 of $g_i \cdots g_{i+k} \times h_j \cdots h_{j+k}$; some of them have appeared earlier and are excluded, the others continue the sequence $\delta(q,i\times j)$. Words of $P\times P$ which have never appeared in this construction never appear in $T^{[ai]_k}w \times T^{[bj]_k}w$ and are just listed after the diamond.

It remains to define the function $\tau: Q \to \{<,>,=\}$ as follows. For a state $q=(s_1t_1\times p_1r_1,\ldots,s_lt_l\times p_lr_l,\diamond,\ldots)$ describing the order of two-letter factors of some $T^{[a]_k}w \times T^{[b]_k}w$ consider the sequence of pairs $c(s_1)\times c(p_1), c(t_1)\times c(r_1), c(s_2)\times c(p_2), c(t_2)\times c(r_2), \ldots, c(s_l)\times c(p_l), c(t_l)\times c(r_l)$ and consider the first of these pairs with non-equal elements, say, $c(t) \gamma c(r)$ with $\gamma \in \{<,>\}$. It indicates the first situation where $T^{[a]_k}v$ and $T^{[b]_k}v$ differ and thus determine the order between the respective numbers. So, $\tau(q)=\gamma$. If such a pair of non-equal elements does not exist, it means precisely that $T^{[a]_k}v=T^{[b]_k}v$. If $a=b$, this is a normal situation, corresponding to $\tau(q)$ equal to $=$. If $a \neq b$ and thus $[a]_k\neq [b]_k$, this means that the sequence $v=c(w)$ is ultimately periodic, and thus the permutation associated with it is not well-defined.

Thus, for each aperiodic automatic word we have constructed an automaton defining the associated permutation. The lemma is proved. \hfill $\Box$

Note that the number of states of the automaton constructed is $O(d^4!)$, where $d$ is the cardinality of the alphabet of the fixed point $w$.  In all the examples we considered, it was possible to obtain an automaton of a much more smaller size: for example, the automaton for the Thue-Morse permutation given above contains only 8 states instead of 16! states of our general construction. However, our method of proof does not allow us to obtain a better general bound.

\section{Non-automatic word generating an automatic permutation}
Suppose that a permutation $\alpha$ is generated by a word $w$ and is $k$-automatic (according to any of the above definitions). Does it imply that the word $w$ is $k$-automatic? The answer to this question is negative.

\begin{example}
{\rm Consider the word $u=u_0u_1 \cdots$ over the alphabet $\{0,1,2\}$ obtained from the Thue-Morse word $w_{TM}=w_0w_1 \cdots$ by substituting some 1s by 2s. More precisely, we write $u_n=2$ instead of $w_n=1$ if and only if the number $.w_nw_{n+1}w_{n+2} \cdots$ is greater than some constant $C$ chosen so that the frequency of 2s in $u$ is irrational. Note that such a constant exists since we can always define the needed irrational frequency as the limit of a sequence of increasing rational frequencies. In all the other cases, we put $u_n=w_n$.

Then clearly $u$ generates the same permutation $\alpha_{TM}$ as the Thue-Morse word since the order between any two entries is preserved under our transformation. In particular, the generated permutation is 2-automatic according to all the three definitions; but the word $u$ is not $k$-automatic for any $k$ since the frequency of $2$  is irrational  (see Th. 8.4.5. from \cite{ash}).}
\end{example}

\section*{Acknowledgements}
An automaton recognizing the permutation generated by an automatic word could be constructed also with a technique due to Allouche, Charlier, Rampersad and Shallit \cite{sh1,sh2}. We are grateful to Prof. J. Shallit for pointing out the references above and for other useful comments.


\begin{thebibliography}{8}
\bibitem{sh1}
J.-P. Allouche, N. Rampersad, J. Shallit, Periodicity, repetitions, and orbits of an automatic sequence. Theor. Comput. Sci. 410 (2009), 2795--2803.
\bibitem{ash}
J.-P. Allouche, J. Shallit, Automatic sequences --- theory, applications, generalizations. Cambridge University Press, 2003.
\bibitem{afks}
S. Avgustinovich, A. Frid, T. Kamae, P. Salimov, Infinite permutations of lowest maximal pattern complexity, http://arxiv.org/abs/0910.5696. Accepted to Theoret. Comput. Sci.
\bibitem{sh2}
\'E. Charlier, N. Rampersad, J. Shallit, Enumeration and Decidable Properties of Automatic Sequences. LNCS v. 6795, Developments in Language Theory 2011, 165--179.
\bibitem{christol}
G. Christol, T. Kamae, M. Mend\`es France, G. Rauzy. Suites alg\'ebriques, automates et substitutions. Bull. Soc. Math. France 108 (1980), 401--419.
\bibitem{cobham}
A. Cobham, Uniform tag sequences. Math. Systems Theory 6 (1972), 164--192.
\bibitem{degs}
J. A. Davis, R. C. Entringer, R. L. Graham, and G. J. Simmons, On permutations
containing no long arithmetic progressions, Acta Arithmetica 34 (1977), 81--90.
\bibitem{eil}
S. Eilenberg, Automata, Languages, and Machines, Vol. A. Academic Press, 1974.
\bibitem{fer}
S. Ferenczi, Complexity of sequences and dynamical systems, Discrete Math.
206 (1999), 145--154.
\bibitem{ff}
D. G. Fon-Der-Flaass, A. E. Frid, On periodicity and low complexity of infinite permutations, European J. Combin. 28 (2007), 2106--2114.
\bibitem{makarov}
M. Makarov, On permutations generated by infinite binary words, Sib. \`Electron. Mat. Izv. 3 (2006), 304--311 [in Russian, English abstract].
\bibitem{mak_pd}
M. Makarov, On the infinite permutation generated by the period doubling word, European J. Combin. 31 (2010), 368--378. 
\bibitem{mak_tm}
M. Makarov, On an infinite permutation similar to the Thue--Morse word, Discrete Math. 309 (2009), 6641--6643.
\bibitem{mak2}
M. Makarov, On the permutations generated by Sturmian words. Sib. Math. J. 50 (2009), 674-–680.


\bibitem{widmer}
S. Widmer, Permutation complexity of the Thue-Morse word, Advances in Applied Mathematics
47 (2011) 309--329.

\end{thebibliography}
\end{document}